# Computer-aided hepatic tumour ablation

David Voirin[1,2], Yohan Payan[1], Miriam Amavizca[1], Antoine Leroy[1], Christian Létoublon[2], Jocelyne Troccaz[1]

[1]Laboratoire TIMC - Faculté de Médecine - Domaine de la Merci
38706 La Tronche cedex - France
[2]Service de Chirurgie générale et digestive - CHU de Grenoble
BP 217 - 38043 Grenoble cedex 9 - France
Author for correspondence: Jocelyne Troccaz,
jocelyne.troccaz@imag.fr

**Abstract.** Surgical resection of hepatic tumours is not always possible. Alternative techniques consist in locally using chemical or physical agents to destroy the tumour and this may be performed percutaneously. It requires a precise localisation of the tumour placement during ablation. Computer-assisted surgery tools may be used in conjunction to these new ablation techniques to improve the therapeutic efficiency whilst benefiting from minimal invasiveness. This communication introduces the principles of a system for computer-assisted hepatic tumour ablation.

## Introduction

Hepatic tumour destructions are traditionally performed under image control (intra-operative CT or intra-operative ultrasounds (US)). The operator introduces the surgical instrument whilst controlling its trajectory on the images. This significantly limits the range of possible trajectories. Our objective is to use a computer-assisted surgery (CAS) approach where pre-operative data are used for planning, intra-operative imaging is limited to registration and the execution of the planned trajectory is guaranteed by the use of a suitable assistance (navigation or robotic systems).

Over the last fifteen years, most of the CAS systems were developed in the context of specialities dealing with bony structures or with structures behaving with limited deformations and movements. Very little attention has been given to computer-assisted soft tissues surgery. Concerning hepatic surgery, the liver position depends both on the patient position and on the point in the respiration cycle. Moreover, hepatic tissues can get intrinsic deformations during surgery (for example, under the action of the surgical instruments). Some recent publications focused on motion evaluation and on registration for CAS liver surgery [1,2]. Meanwhile, registration experiments were performed on static phantoms. One purpose of this preliminary work was to quantitatively evaluate the feasibility of registration. The image modalities that were chosen for this evaluation were the ones that are traditionally selected for registration, namely pre-operative CT or MRI and intra-operative ultrasonic imaging.

## Experiments

The main objective was to be able to quantitatively evaluate the algorithms that match *pre-operative data (CT or MRI) with 2.5D echography*[1] under rather realistic conditions. In the first stage of this study, it was assumed that the key steps of the protocol, namely image acquisitions and guidance, can be both executed at a same point in the respiration cycle, for instance after expiration. Because the experiments were conducted with a volunteer, MRI was preferred to CT. The acquisition was synchronized with the respiration cycle and occurred only during the expiration phase. In a second stage - that simulates intra-operative procedures - 2.5D echographic acquisitions were performed. Data were acquired during the apnoea that follows expiration. Based on these data, the registration algorithm - a surface rigid matching using a distance map recorded in an octree-spline data structure [3] - was quantitatively evaluated in terms of repeatability and accuracy.

Data registration was repeatable and the accuracy - even limited by data acquisition - allows envisioning the clinical applicability of the method. A detailed presentation and discussion of these results may be found in [4].

## Acknowledgements

This work has been supported by La Fondation de la Recherche Médicale. We thank Patrick Vassal, Pr Lebas, Pr Coulomb, Dr Sengel and Dr Teil for their assistance in the image acquisition processes.

---

[1] 2.5D echography is the process where the ultrasonic probe is localized in space thanks to a localizer (Optotrak in this experiment) allowing to get sparse 3D data.